\documentstyle[12pt]{article}

\newcommand{\add}{\addtocounter{eqncnt}{1}}
\newcounter{eqncnt}[section]

\newcommand{\be}{\begin{equation}}
\newcommand{\ee}{\end{equation}\add}
\newcommand{\bea}{\begin{eqnarray}}
\newcommand{\eea}{\end{eqnarray}}

\newcommand{\ol}{\overline}

\newcommand{\bra}{\left\langle}
\newcommand{\ket}{\right\rangle}

\newcommand{\mpc}{\mbox{$h^{-1}$~Mpc}}

\begin{document}
\begin{center}
{\Large \bf Null geodesics and observational cosmology} \\[2mm]

\vskip .5in

{\sc A.A. Coley}\\
{\it Department of Mathematics and Statistics}\\
{\it Dalhousie University, Halifax, NS B3H 3J5, Canada}\\
{\it aac@mathstat.dal.ca }
\end{center}

\begin{abstract}

The Universe is not isotropic or spatially homogeneous on local
scales. The averaging of local inhomogeneities in general
relativity can lead to significant dynamical effects on the
evolution of the Universe, and even if the effects are at the 1\%
level they must be taken into account in a proper interpretation
of cosmological observations. We discuss the effects that
averaging (and inhomogeneities in general) can have on the
dynamical evolution of the Universe and the interpretation of
cosmological data. All deductions about cosmology are based on the
paths of photons. We discuss some qualitative aspects of the
motion of photons in an averaged geometry, particularly within the
context of the luminosity distance-redshift relation in the simple
case of spherical symmetry.

\end{abstract}

\newpage


\newpage

\section{Introduction}

The averaging problem in cosmology is perhaps the most important
unsolved problem in mathematical cosmology.  The averaging of
local inhomogeneities in general relativity (GR) can lead to very
significant dynamical effects on the evolution of the Universe. We
discuss the effects that averaging (and inhomogeneities in
general) can have on the interpretation of cosmological data. All
deductions about cosmology are based on the paths of photons. In
particular, we discuss some theoretical aspects of the motion of
photons in an averaged geometry.

Cosmological observations \cite{Riess:2006fw,bennett}, based on
the assumption of a spatially homogeneous and isotropic
Friedmann-Lema\^{i}tre-Robertson-Walker (FLRW) model plus small
perturbations, are usually interpreted as implying that there
exists dark energy, the spatial geometry is flat, and that there
is currently an accelerated expansion giving rise to the so-called
$\Lambda$CDM-concordance model with $\Omega_m \sim 1/3$ and
$\Omega_{de} \sim 2/3$. Although the concordance model is quite
remarkable, it does not convincingly fit all data
\cite{bennett,probs,Sarkar,Kash}. Unfortunately, if the underlying
cosmological model is not a perturbation of an exact flat FLRW
solution, the conventional data analysis and their interpretation
is not necessarily valid. There is the related question of whether
the Universe is `exactly FLRW' on some large scale (the scale of
homogeneity is presumably $\sim 400$ $h^{-1}$Mpc or greater, since
there is strong evidence for coherent bulk flows on scales out to
at least 300$h^{-1}$Mpc \cite{Kash}), or only approximately FLRW
at all scales (in which case there is also the `fitting problem'
of which background FLRW model to take \cite{ELLIS}).

Inhomogeneities can affect observational calculations
\cite{Wiltshire}. Supernovae data (in combination with other
observational data), dynamically requires an accelerating
universe.  However, this only implies the existence of dark matter
if the Universe  is (approximately) FLRW. Supernovae data can be
explained without dark matter in inhomogeneous models, where the
full effects of GR come into play. The
apparent acceleration of the Universe is thus not caused by
repulsive gravity due to dark energy, but rather is a dynamical
result of inhomogeneities, either in an exact solution or via
averaging effects. In an exact inhomogeneous cosmological model,
the inhomogeneities affect the dynamics of the model, and a
special observer can indeed measure the observed effects,
although the results are, of course, model  dependent.

In particular, averaging effects are real, but their relative
importance must be determined.  In the  FLRW plus perturbations
approach, the (backreaction) effects are assumed small (and are
assumed to stay small during the evolution of the universe). A
perturbation scheme is chosen, and assumptions are made so that
the approximation scheme is well defined (i.e., assumptions are
made so that there are no divergences and the backreaction stays
small at the level of the perturbation, and the estimated
corrections are small by definition \cite{BRB}). This is perhaps a
cyclical argument, but is has merit since it suggests that there
is a self-consistent argument that backreactions are and remain
small within the FLRW plus perturbations approach. Presumably,
divergences would signal a breakdown in the approximation scheme
when backreaction effects could be large.

There is also the possibility that averaging effects are not small
(i.e., perturbation theory cannot be used to estimate the effects,
and real inhomogeneous effects must be included). This is much
more difficult to study, but the results from exact inhomogeneous
solutions perhaps suggest that this ought to be investigated
properly. There is also the question of what it means for an
effect not to be small (and hence not negligible). It has been
argued that an affect at the 1\% level can be regarded as significant,
and could very well affect the dynamics and interpretation of
observational data \cite{Col}.

{\em{Averaging}}: The averaging problem in cosmology is of
considerable importance for the correct interpretation of
cosmological data. The correct governing equations on cosmological
scales are obtained by averaging the Einstein equations. By
assuming spatial homogeneity and isotropy on the largest scales, the
inhomogeneities affect the dynamics though correction
(backreaction) terms, which can lead to behaviour qualitatively
and quantitatively different from the FLRW models.

The macroscopic gravity (MG) approach is an exact approach which
gives a prescription for the correlation functions that emerge in
an averaging of the Einstein field equations \cite{Zala}. In
\cite{CPZ} the MG equations were explicitly solved in a FLRW
background geometry and it was found that the correlation tensor
(backreaction) is of the form of a spatial curvature. This result
was confirmed in subsequent work in which  the spherically
symmetric Einstein equations were explicitly averaged
\cite{CPpaper}, and is consistent with the scalar averaging
approach of Buchert \cite{Buchert:1999pq} in the Newtonian limit
and with the results of averaging an exact
Lema\^{\i}tre-Tolman-Bondi (LTB) model and the results of linear
perturbation theory \cite{Bild-Futa:1991}. There is no question
that the backreaction effect is real. The only question remaining
is the size of the effect. However, even a small non-zero
curvature $|\Omega_{k}|\sim .01$ would lead to significant effects
on observations (for redshifts $z \geq 0.9$) \cite{dunkley};
indeed, values of $|\Omega_{k}| \sim 0.05$ or larger have been
found to be consistent with observation \cite{IT06}.

{\em{Curvature estimates}}: The Wilkinson Microwave Anisotropy
Probe (WMAP) has  reported $\Omega_{tot} = 1.02 \pm 0.02$
\cite{bennett}. In $\Lambda$CDM models type Ia supernovae (SNIa)
data alone  prefers a slightly closed Universe
\cite{Riess:2006fw}. Taken at face value this suggests $\Omega_{k}
= 0.02$. Models with non-negligible curvature have been studied
recently \cite{IT06}. Indeed, models with no dark energy have been
found to be consistent with supernovae data and WMAP data
\cite{Sarkar}. Combining these observations with Large Scale
Structure (LSS) observations such as the Baryon Acoustic
Oscillations (BAO) data put stringent limits on the curvature
parameter in the context of adiabatic $\Lambda$CDM models.
However, these data analyses are very model- and prior-dependent
\cite{Shapiro:2005}, and care is needed in the proper
interpretation of data.

Let us discuss the observational estimates in more detail. To
first order, the averaged curvature $\Omega_{k}$ evolves like a
constant--curvature model \cite {NewLiS,Rasanen:2003fy,Col}. In
the course of expansion the second order kinematic backreaction
$\Omega_{\cal Q}$  becomes dynamically more important and can
become significant when the structure formation process injects
backreaction \cite{NewRas}. For typical scales $\ell \sim$ 50 -
200 \mpc ~and for a Hubble radius $\ell_H\sim$ 3000 \mpc, values
of second order effects $\sim (\ell/\ell_{H})^2$ are typically in
the range $0.01$ to $0.1$. Robust estimators for intrinsic
curvature fluctuations using realistically modelled clusters and
voids in a Swiss-cheese model indicates that the dark energy
effects can be reduced by up to 30 \%
\cite{buch,hellaby:volumematching}. A rough order of magnitude
estimate for the variance implied by the observed density
distribution of voids implies $|\Omega_{\cal Q}|\sim 0.1-0.2$
\cite{NewRas}.  The backreaction parameter has also been estimated
in the framework of Newtonian cosmology \cite{Buchert:1999pq}; it
was found that the backreaction term can be quantitatively small
(e.g., $\Omega_{\cal Q}=0.01$), but the dynamical influence of a
non--vanishing backreaction,
$\Omega_{I} = \Omega_{k} + \Omega_{\cal Q}$,  
on the other cosmological parameters can, in principle, be
quite large.

From perturbation theory, it might be expected that the
backreaction effect $\Omega_{I}\sim 10^{-5}-10^{-4}$
\cite{NewLiS,BRB} (naively it is expected that the order of the
pertubation, consistent with CMB data, is $\sim 10^{-5}$
\cite{EGS}). From inhomogeneous models larger effects
might be expected. There is a heuristic argument that
$\Omega_{I}\sim 10^{-3}-10^{-2}$ \cite{Rasanen:2003fy,Col,BRB},
which is consistent with CMB observations \cite{Col}. Let
$\Omega_{k} \sim \epsilon$. We can estimate ${\cal Q} \sim \bra
{\sigma^2}\ket_D$, where ${\sigma_D}$ is the fluctuation
amplitude; then $\Omega_{\cal Q}\sim \bra {\sigma^2}\ket_D/H^2_D
\sim \bra {\delta}\ket_D \sim \epsilon^2$ (where $\delta$ is the
density contrast) \cite{NewLiS}. Consequently, $|\Omega_{k}| \sim
10^{-2}$. Indeed, from perturbation theory the dominant local
corrections to redshift or luminosity distance go as $\nabla
\Phi$, which is only suppressed as $\ell/\ell_H$ (rather than as
the Newtonian potential $\Phi$ as naively expected, which is
suppressed as $(\ell/\ell_H)^2)$ \cite{Rasanen:2003fy}; for scales
$\sim$ 100 \mpc, $\Omega_{\cal Q} \sim 10^{-5}$ \cite{NewRas}. A
possible scenario in which $\Omega_{I} \sim 0.2-0.4$, which can
occur in exact inhomogeneous models  (for example, many authors
have studied observations in LTB models, albeit with contradictory
conclusions \cite{LTBgeo}) and in which it is possible for the
observed acceleration to be explained (by backreaction) using the
standard interpretation without resorting to dark energy, is
probably not supported by most authors \cite{buch,NewRas}.

This consequently suggests a scenario in which $|\Omega_{I}|\sim
0.01-0.02$. It must be appreciated that this value for
$\Omega_{I}$ is relatively large (e.g., it is comparable with the
total contribution of baryons to the normalized density) and may
have a significant dynamical effect.  Perhaps cosmological data
can be explained without dark energy through a small backreaction
and a reinterpretation of cosmological observations
(alternatively, dark energy may still be needed for consistency
with observations). However, a value of $|\Omega_{I}| \sim
0.01-0.02$ would certainly necessitate a complete reinvestigation
of cosmological observations. In addition, such a value cannot be
explained by inflation. From standard inflationary analysis,
$|\Omega_{tot} - 1|$ would be effectively zero, so that any
non-zero residual curvature might only be naturally explained in
terms of a backreaction. Let us consider this in more detail.

{\em{Observations}}: Clearly, averaging can have a very
significant dynamical effect on the evolution of the Universe; the
correction terms change the interpretation of observations so that
they need to be accounted for carefully to determine if a model
may be consistent with cosmological data. Averaging may or may not
explain the observed acceleration. However, its effects cannot be
neglected. This leads us to a paradigm shift and clearly
cosmological observations need to be revisited. Any observation
that is based on physics in the nonlinear regime may well be
influenced by the backreaction effect.

The standard analysis of SNIa  and CMB data in FLRW models cannot
be applied directly when backreaction effects are present, because
of the different behaviour of the spatial curvature
\cite{Shapiro:2005}. Indeed, in principle all data needs to be analysed within
a particular inhomogeneous model, and not just an averaged
version. Studies of the LTB model have demonstrated that the
effect of inhomogeneity on the luminosity distance can mimic
acceleration \cite{LTBgeo}.

It is necessary to carefully identify observables actually
measured by an experiment. For example, there are several
different measures of the expansion rate and acceleration. In
addition, there is the question of whether we are dealing with
regional dynamics or global (averaged) dynamics  in any particular
observation. The determination of the Hubble parameter and SNIa
observations (which need data at low redshift $z<1$) are clearly
based on local measurements. However, most observations of the CMB
(e.g., the integrated Sachs-Wolfe effect, which can be probed at
large redshifts) and LSS are sensitive to large-scale (averaged)
properties of the Universe.

Observations can clearly allow for a non-zero curvature
\cite{Sarkar,IT06,Shapiro:2005}. In principle, in the new paradigm
parameters can be implicit functions of the spatial scale; in
particular, the curvature parameter can  be different for
different averaging domains. Thus, it is of interest to
investigate the possibility of a scale dependent spatial
curvature. Different observations might probe effective curvatures
at different scales. In fact, SNIa data alone do not efficiently constrain
curvature and the addition of other cosmological
probes are necessary (at both small and large scales). It is important
to test the idea of whether different quantities (valid
at different scales) are measured in different experiments. In
particular, the spatial curvature in SNIa experiments (small
scale, $z<<2$) might be  a different spatial curvature to that
measured in CMB experiments (large scale, $z \sim 1100$). We could
parameterize the different observations with different effective
observables corresponding to different effective parameters (e.g.,
$\Omega_k$ for SNIa  and $\Omega_{\ol{k}}$ for CMB). At first read,
data from CMB and SNIa support $|\Omega_{\ol{k}}-\Omega_k| \sim
0.02$; i.e., the global (average) and local values of the
curvature constant  can differ by up to about 2\%; that is, SNIa
and CMB data allow the possibility of scale dependent spatial
curvatures. It might be useful to do a  comprehensive statistical
analysis of cosmological data and include different scale
dependent spatial curvatures as parameters (`parameter splitting') to be fitted to find
their best fit values.

\newpage

\section{Null geodesics}

All deductions about cosmology are based on light paths. The
present assumption is that intervening inhomogeneities average
out. However, inhomogeneities affect curved null geodesics
\cite{NewRas,buch} and can drastically alter observed distances
when they are sizable fraction of the curvature radius. In the
real Universe, voids occupy a much larger region as compared to
structures \cite{Hoyle:2003}, hence light preferentially travels
much more through underdense regions and the effects of
inhomogeneities on luminosity distance are likely to be
significant. Spherically symmetric models are a useful testing
ground for studying the possible effects of inhomogeneities on
cosmological observations. It has been shown that in spherically
symmetric (LTB) spacetime model spatial variations of the
expansion rate can significantly affect the cosmological
observations (particularly close to the center) of a spherically
symmetric model \cite{LTBgeo}.

Let us discuss the effect of averaging on null geodesics in
inhomogeneous models. We treat GR as a microscopic (classical)
theory \cite{TavZal}. In this idealization real particles are
modelled as point particles moving along timelike geodesics in the
absence of external forces. The physical quantities which are
observables must be identified. However, because all observations
are of finite resolution, this involves a finite region of
spacetime and therefore observations are extended (in the sense
that they necessarily involve averages of measured quantities). In
practice, only spacetime averages of field quantities have
physical meaning; real observations in cosmology are invariably
extended (in the sense of covering a neighbourhood of spacetime),
averaging over spacetime of some characteristic length (not the
same as cosmological scale though). Thus GR as a microscopic
(classical) theory is incomplete and only an approximation.

The question arises of how to interpret observations in an
inhomogeneous universe \cite{LTB2}.
Only the redshift and the energy flux of light arriving from a
distant source are observed, rather than expansion rate or matter
density, and we need to know how to relate actual observations to
the various quantities in the equations (either averaged or
non-averaged). We need to {\em define} a distance scale function
and the corresponding Hubble rate; in inhomogeneous spacetime
there will be different possible definitions and they will, in
general, depend on both space and time.

Studying  observations in inhomogeneous spacetimes is hard. Here
we shall just discuss some qualitative aspects within the context
of the observed luminosity distance-redshift relation.
Investigating the CMB is much more  difficult; in order to compute
the spectrum of temperature anisotropies a full perturbation
theory (e.g., in a spherically symmetric spacetime) is needed.

One problem of interpreting observations is that it is necessary,
in principle, to model properties of (not only a single photon)
but of a `narrow' beam of photons. There is also the problem of
which observations deal with macro-'geometric averages' and which
deal with statistical averages (of single observed micro-values);
i.e., the question of spacetime averaging verses statistical
properties. This cannot be done adequately within the Buchert \cite{buch}
approach to the averaging problem, where only scalar quantities
are averaged.

Real observations involve a beam or bundle of photons (a local congruence of null
geodesics). From the geometric optics approximation for a test
electromagnetic field we can obtain the optical scalar equations
that govern the propagation of the shearing and expansion (of the
cross-sectional area of the beam) with respect to the affine
parameter along the congruence due to Ricci focussing and Weyl
tidal focussing \cite{Dyer}. In FLRW models (macro-geometry),
there is only Ricci focussing (the Weyl tensor is zero). However,
in reality photons from distant sources have passed around massive
objects along the line of sight and locally have only
experienced Weyl focussing in the micro-geometry. Since the
optical scalar equations (which are {\em nonlinear}) require
integration along the beam, the optics for a lumpy distribution
does not average and there may be important resulting effects. So
it is also of importance to study the effect of averaging on a
beam of photons in the optical limit. In addition, due to the
positive-definite nature of shear focussing, shearing can cause
local refocussing of the beam leading to the formation of caustics
(which are not present in the averaged FLRW macro-geometry), which
can lead to further observational effects but which are not of primary
interest here.

In addition to the question of what are the observations involved
actually measuring and statistical averages verses geometrical
averages, there is the question of the motion of photons in an
averaged geometry. Are photons moving on null geodesics of the
(averaged) macro-geometry? Are observations dealing with an
(extended) beam of photons? If so, how does this relate to the
motion of a beam of photons in micro/macro geometry. We shall
illustrate these issues with a simple example of radial null
geodesics in spherical symmetry within MG.

{\em{Null geodesics in MG}}: The microscopic field is to be
averaged on an intermediate spatial scale, $\ell$ (which is, in
principle, a free parameter of the theory, and assumed large
compared to the scale on which astrophysical objects such as
galaxies or clusters of galaxies have structure and is usually
tacitly assumed to be a few hundred Mpc, or a fraction of the
order of the inverse Hubble scale, $\ell_H$). We then obtain a
macro- (averaged) geometry (using, for example, non-perturbative
MG).

Let us consider the motion of photons in an averaged geometry, and
investigate the resulting effect on cosmological observations. We
assume GR is a microscopic theory on small scales, with local
metric field ${\bf g}$ (the geometry) and matter fields.  A photon
follows a null geodesic ${\bf k}$ in the local geometry.  After
averaging we obtain a smoothed out macroscopic geometry (with
macroscopic metric $\overline{{\bf g}} = \langle {\bf g} \rangle$)
and macroscopic matter fields, valid on larger scales. But what
trajectories do photons follow in the macro-geometry.  Is the
``averaged'' vector $\langle {\bf k} \rangle \equiv \overline{{\bf k}}$
null, geodesic (or affinely parametrized) in the macro-geometry.
How does this affect cosmological observations?

Indeed, in the micro-geometry
\begin{eqnarray}
g_{ab} k^a k^b  & = & 0  \mbox{  (null) }\\
\bigtriangledown k^a \equiv k^a_{~;b} k^b & = & 0 \mbox{
(geodesic; affinely parametrized)}.
\end{eqnarray}

After averaging, in general
\begin{eqnarray}
\langle g_{ab} k^a k^b \rangle & \neq & \langle g_{ab}\rangle
\langle k^a \rangle \langle k^b\rangle = \overline{g}_{ab} \overline{k}^a
\overline{k}^b \\
\label{Bb} \overline{\bigtriangledown} ~\overline{k}^a =
\overline{k}^a_{~|b}\overline{k}^b & \neq & 0        ,
\end{eqnarray}
where covariant differentiation $(\overline{\bigtriangledown},|)$ is with
respect to the
macroscopic metric.

There is also the question of how to define the averaged matter,
and whether (for example) the averaged matter moves on timelike
geodesics (by the same reasoning the average of a timelike vector
need not be timelike). This is not directly relevant here. In
principle, it is possible to study the effects of averaging on the
motion of the matter fields from the microphysics (e.g., from
kinetic theory). However, this is part of the assumptions of the
matter in the (macro) cosmological model and is, as in micro-GR,
an ``external'' assumption on the matter model (and these
assumptions can be probed separately).  If the macro-matter is
assumed to be pressure-free dust, then from the conservation
equations the dust will move on timelike geodesics of the averaged
macro spacetime.

As an illustration, let us consider these questions in a spherically symmetric
spacetime given in volume
preserving coordinates (VPC), which enables us to calculate the averaged
quantities in a relatively straightforward manner. The
(non-comoving) spherically symmetric metric  in local VPC
($\sqrt{-g}=1$, where $g=det(g_{ab})$
\cite{CPpaper}), is given by

\begin{equation}
ds^2=-Bdt^2+Adr^2+\frac{1}{\sqrt{AB}}d\Sigma^2, \label{vpcss}
\end{equation}

\noindent where the functions $A$ and $B$ depend on $t$ and $r$.

Let us consider an incoming radial null ray given by
\begin{eqnarray}
{\bf k} = K\left( \frac{\partial}{\partial t} -
\sqrt{\frac{B}{A}} \frac{\partial}{\partial r}  \right).
\end{eqnarray}
A direct calculation yields
\begin{eqnarray}
k^a_{~;b}k^b = G^a = Gk^a,
\end{eqnarray}
where \be G \equiv K \left\{ \frac{K_t}{K} + \sqrt{\frac{B}{A}}
\frac{K_r}{K} + \left[ \frac{1}{2} \frac{B_t}{B} + \frac{1}{2}
\frac{A_t}{A} + \sqrt{\frac{B}{A}} \frac{B_r}{B}  \right]   \right
\}. \ee Here $k^a$ is automatically a null geodesic; it is
affinely parametrized with affine parameter $\lambda$, if $G = 0$.

The averaged metric is given by \be ds^2 = -\overline{B} dt^2 +
\overline{A} dr^2 + d \overline{\Sigma}^2, \ee and we write \be
\overline{\bf k} = \overline{K} \frac{\partial}{\partial t} -
\overline{L} \frac{\partial}{\partial r}. \ee We see, in general,
that from (\ref{Bb}). \be \overline{g}_{ab} \overline{k}^a
\overline{k}^b \neq 0; \ee indeed, $\overline{k}^a$ is null only
when

\be \label{7} \overline{A} \; \overline{L}^2 = \overline{A}
\left\langle K \sqrt{\frac{B}{A}}\right\rangle^2 =
\overline{B}~\overline{K}^2. \ee Calculating further, we obtain
\be \overline{\bigtriangledown} \; \overline{k}^a \equiv \overline{H}^0
\frac{\partial}{\partial t} + \overline{H}^1
\frac{\partial}{\partial r}, \ee where \bea \overline{H}^0 & =
&(\ol{K}_t \ol{K} + \ol{K}_r \ol{L}) + \frac{1}{2\ol{B}} (\ol{B}_t
\ol{K}^2 + 2\ol{B}_r
\ol{K} \; \ol{L} + \ol{A}_t \ol{L}^2)\nonumber \\
&& \\
\ol{H}^1 & = & (\ol{L}_t \ol{K} + \ol{L}_r \ol{L}) + \frac{1}{2
\ol{A}} (2\ol{A}_t \ol{L} \;\ol{K} + \ol{A}_r \ol{L}^2 + \ol{B}_r
\ol{K}^2) \nonumber \eea Clearly $\ol{K}^a$ is not necessarily
geodesic.

Assuming that $\ol{k}^a$ is null, only if $\ol{H}^1 = -
\sqrt{\frac{\ol{B}}{\ol{A}}} \ol{H}^0$, is $\ol{k}^a$ then
geodesic. However, even then it is certainly not affinely
parametrized (the equivalent ``averaged'' quantity $\ol{G}$ is not
zero, and $\ol{\lambda}$ is not an affine parameter).
[Alternatively, presumably we could define a new ``averaged'' geodesic null
vector (so that equation (\ref{7}), for example, is satisfied),
but it is not necessarily affinely parametrized (and it is not clear whether such
a quantity is related to observations).]

The form of the correlation tensor now depends on the assumed form
for the inhomogeneous gravitational field and matter distribution.
Let us assume  \cite{CPpaper}\bea
A(r,t) & = \langle A(r,t) \rangle [1+ L_0 a(t) \delta_a(r,t) + {\mathcal{O}}(L_0^2)]\nonumber\\
&& \\
B(r,t) & = \langle B(r,t) \rangle [1+ L_0 b(t) \delta_b(r,t) +
{\mathcal{O}}(L_0^2)]\nonumber \eea where $L _0= \ell/\ell_H \sim 10^{-1}<1$
and $ \langle \delta_i(r,t) \rangle = 0$. From \cite{CPpaper}, the
correlation (correction) tensor in the case of a FLRW
macro-geometry is of the form of a spatial curvature term with
parameter $k_0$ (where $k_0$ is an integration constant depending
on the actual averaging and, in particular, the averaging scale;
i.e., it is not the spatial curvature parameter of the
micro-geometry) [in more generality, the correlation tensor has an
additional anisotropic contribution (correction), which is absent in
the FLRW macro-geometry \cite{CPpaper}]. In this case $\ol{G}^1 =
- \sqrt{\frac{\ol{B}}{\ol{A}}} \ol{G}^0 + {\mathcal{O}}(L_0)$, and the affine
parameter is given by \be \ol{\lambda} \simeq \lambda(1+ k_0
\delta(t)). \ee

Observational quantities, in terms of the luminosity distance
${d}_L$ and the redshift $z$ for example, can be determined from
the underlying geometry \cite{LTB2}. For a flat FLRW background,
the luminosity distance - redshift relation then becomes (for
small redshift) \be \ol{H}_0 \ol{d}_L \simeq \ol{z} + \frac{1}{2}
(1- \ol{q}_0 + C) \ol{z}^2 +  {\mathcal{O}}(\ol{z}^3),\ee where $C \simeq k_0
\epsilon(t)$. [Additional corrections, due to the fact that the
trajectories are not exactly null geodesics and possible
anisotropic averaging corrections (contained in $\ol{q}_0$) might
be expected to be small ($\sim {\mathcal{O}}(L)$)--the exact values would
depend on the averaging scheme, the assumed form of
inhomogeneities, the averaging scale, etc.).]

That is, the averaging effectively renormalizes $\ol{q}_0$, and
will include an additional contribution, $C$, from the ``spatial
curvature'' with parameter $k_0$ that occurs from the averaging.
This type of effect, $(1- {q}_0) \rightarrow (1- \ol{q}_0 + C)$,
where $C$ is a dimensionless function of time characterizing the
possible inhomogeneities (and perhaps depending on the density and
pressure of the matter and age and expansion of the Universe), and
its potential affect on cosmological observations has, in fact,
been considered previously (and can be important, depending on
which observations are considered) \cite{PM}.

{\em{Discussion}}: It is clear there are a number of effects, all
of order of about 1\%, which will add up and perhaps produce a
significant observational effect. There is a  1\% effect due to the non-affine parameterization
of the null geodesics in the averaged geometry, as illustrated above, together with  
additional corrections due to the fact that the
trajectories are not exactly null geodesics, which are expected to be of a
similar order of magnitude. There are  corrections due to the fact that
the correlation tensor is not precisely due to a spatial curvature 
(and might possibly contain anisotropic corrections). There are also corrections 
due to the fact
that a beam of photons will experience expansion and shearing (and perhaps even twisting)
during their evolution  \cite{Dyer}; these effects are again expected to
be at the level of about one percent \cite{muns}. There is also the effect of a special location
of observers in an exact inhomogeneous cosmological model; although such an
effect would influence all observations, it might have a different quantitative effect
on small scale observations (in the non-linear regime where local inhomogeneities 
have a more significant effect) than 
on large scale observations like the CMB (where linear effects might effectively cancel
when integrating over scales larger than the homogeneity scale).
Clearly, averaging and non-zero spatial curvature can have
important effects, especially on SNIa measurements, and cannot be
neglected \cite{clarkson}.

For example, using cosmological observations to determine the equation of state
of dark energy is unreliable, since this is exactly where  the
effects of averaging are expected to be important. Within dark energy
models, with an effective equation of state parameter $w$,
parameterized (for example) by $w(a) = w_0 + (1-a)w_a $\cite{CPL},
there has been an attempt to probe $w_a$. However, the effect of spatial
curvature/averaging which have been neglected (and which are as much as a 1\% effect
\cite{clarkson,Col}), is as large as the effects being probed.

\newpage

{\em Acknowledgements}. I would like to thank Rob Crittenden, Robert van den Hoogen,
Elizabetta Majerotto, Nicos Pelavas,
Bill Stoeger and Roberto Sussman  for
helpful discussions.


\begin{thebibliography}{abc}

\baselineskip 12pt


\bibitem{Riess:2006fw} P. Astier {\it et al}, Astron. Astrophys. {\bf 447}, 31
(2006); T.M. Davis {\it et al}, Ap. J. {\bf 666}, 716 (2007).




\bibitem{bennett} J. Dunkley {\em et al}, arXiv:0803.0586; D. N. Spergel {\it et al}, Ap. J. Suppl. {\bf 170},
377 (2007) [astro-ph/0603449].



\bibitem{probs} S. Ho {\em et al}, Phys. Rev. D
{\bf 78} 043519 (2008); C. L. Reichardt {\em et al},
arXiv:0801.1491; M. L. McClure and C. C. Dyer, New. Astron. {\bf
12} 533 (2007); U. Seljak {\em et al}, JCAP {\bf 0610} 014 (2006).




\bibitem{Sarkar} A. Blanchard, {\em et al},
Astron. Astr.  {\bf 449} 925 (2006).

\bibitem{Kash} A. Kashlinsky {\em et al.}, arXiv:0809.3734



\bibitem{ELLIS} G. F. R. Ellis {\em et al.}, Phys. Rep. {\bf 124} 315
(1985); G. F. R. Ellis and W. Stoeger, Class. Quant. Grav. {\bf 4}
1697 (1987).

\bibitem{Wiltshire} D.L.Wiltshire, Phys. Rev. Lett. {\bf 99} 251101 (2007) \&
Phys. Rev. {\bf D 78} 084032 (2008).

\bibitem{clarkson} C. Clarkson  {\em et al.}, JCAP {\bf 0708} 011 (2007).

\bibitem{BRB} A. Paranjape and T. P. Singh, Phys. Rev. Letts. {\bf 101}
(2008); A. Paranjape, Int. J. Mod. Phys. D {\bf 17} 597 (2008); I.
A. Brown, G. Robbers and J. Behrend, [arXiv:0811.4494].

\bibitem{Col} A. Coley, [arXiv:0704.1734]


\bibitem{Zala} R. M. Zalaletdinov, Gen. Rel. Grav. {\bf 24}
1015 (1992) \& {\bf 25} 673 (1993).

\bibitem{CPZ} A. A. Coley {\em et al},
Phys. Rev. Letts. {\bf 595} 115102  (2005) [gr-qc/0504115].


\bibitem{CPpaper} A. A. Coley and N. Pelavas,
Phys. Rev. D {\bf 75} 043506 (2006) \& {\bf 74} 087301 (2006).

\bibitem{Buchert:1999pq}
T. Buchert, M. Kerscher and C. Sicka, Phys. Rev. D {\bf 62} 043525
(2000).


\bibitem{Bild-Futa:1991} T. Futamase,  Phys. Rev. D {\bf 53}
681 (1993) \& Prog. Theor. Phys. {\bf 9} 58 (1993).



\bibitem{dunkley}
J. Dunkley {\em et al}, Phys. Rev. Lett. {\bf 95} 261303 (2005).

\bibitem{IT06} E.~Wright,
astro-ph/0701584; R. Aurich {\em et al}, Class. Q. Grav. {\bf 22}
3443 (2005).

\bibitem{Shapiro:2005} C. Shapiro and M. S. Turner, Ap. J. {\bf 649} 563 (2006)
[astro-ph/0512586].

\bibitem{NewLiS} N. Li and D. J. Schwarz, Phys. Rev. D {\bf 78} 083531 (2008).


\bibitem{buch} T. Buchert. Gen. Rel. Grav.
{\bf 33} 1381 (2001) \& {\bf 32} 105 (2000).

\bibitem{NewRas} S. R\"{a}s\"{a}nen, JCAP {\bf 11} 003 (2006) \& .JCAP {\bf 0804} 026 (2008)


\bibitem{hellaby:volumematching}
C. Hellaby, Gen. Rel. Grav. {\bf 20}, 1203 (1988).

\bibitem{EGS} R. Maartens {\em et al},  Phys. Rev.  D {\bf 51}
1525 \& 5942 (1995).


\bibitem{Rasanen:2003fy} S. R\"{a}s\"{a}nen, JCAP {\bf 02} 003 (2004).

\bibitem{LTBgeo} J. Garcia-Bellido and T. Haugbolle,
arXiv:0802.1523; H. Alnes {\em et al}  Phys. Rev. D. {\bf 73}
083519 (2006); K. Enqvist and T. Mattsson, JCAP {\bf 0702} 19
(2007); E. Barausse {\em et al},  Phys. Rev. D. {\bf 71} 063537
(2005).


\bibitem{Hoyle:2003} F. Hoyle and M. S. Vogeley ,
Ap. J. {\bf 607} 751 (2004).

\bibitem{TavZal} R. Tavakol and R. Zalaletdinov, Found. Phys. {\bf 28} 307 (1998).

\bibitem{FTH} J.A. Frieman, M.S. Turner and D. Huterer,
arXiv:0803.0982.

\bibitem{Dyer} R.K. Sachs, Proc. Roy. Soc. (London)  {\bf A264} 309
(1974); C.C. Dyer, Gravitational Lenses and the Inhomogeneous
Universe, in Theory and Observational Limits in Cosmology (Proc.
Vatican Observatory Conference at Castel Gondalfo), ed. W.R.
Stoeger (Vatican Observatory Press, 1985); C.C. Dyer and R. C.
Roeder, Ap. J. {\bf 189} 167 (1974).

\bibitem{LTB2} G.R.F. Ellis, General Relativity and Cosmology,
in Proc. Int. School of Phys., Enrico Fermi, ed. B.K. Sachs
(Academic Press, 1971); R. Maartens, N. Humphreys and D.
Matravers, arXiv: gr-qc/9511045; N. Mustapha, C. Hellaby and
G.R.F. Ellis, MNRAS {\bf 292} 817 (1997).


\bibitem{PM} M.H. Partori and G. Mashoon, Ap. J. {\bf 276} 4
(1984).


\bibitem{CPL} E.V. Linder, Phys. Rev. Lett. {\bf 90} 091301 (2003).


\bibitem{muns} D. Munshi {\em et al.}, Phys. Rep. {\bf 462} 67
(2008).



\end{thebibliography}
\end{document}